\documentclass[aps,prd,preprint,preprintnumbers]{revtex4}
\usepackage[utf8]{inputenc}
\usepackage{slashed}
\usepackage{fancybox}
\usepackage{hhline}
\usepackage{dcolumn}
\usepackage{textcomp}
\usepackage{epsfig,graphics,graphicx}
\usepackage{amsfonts,amssymb,amsmath}
\usepackage{pifont}
\usepackage{bm}
\usepackage{longtable} 
\usepackage{appendix}
\usepackage[mathscr]{euscript}
\usepackage{mathrsfs}
\usepackage{multirow}
\usepackage{rotating}
\usepackage{color}   

\newcommand{\sh}[1]{\slash{\!\!\!\! #1}\;}
\newcommand{\beq}{\begin{equation}}
\newcommand{\eeq}{\end{equation}}
\newcommand{\bea}{\begin{eqnarray}}
\newcommand{\eea}{\end{eqnarray}}

\newcommand{\eps}{\epsilon}
\newcommand{\ord}[1]{{\cal{O}}( #1 )}

\DeclareFontFamily{OT1}{pzc}{}
\DeclareFontShape{OT1}{pzc}{m}{it} 
{<-> s * [0.900] pzcmi7t}{}
\DeclareMathAlphabet{\mathpzc}{OT1}{pzc} 
{m}{it}
\DeclareMathAlphabet{\mathcalligra}{T1}{calligra}{m}{n}
\usepackage{hyperref}
\hypersetup{pdfauthor={me}, colorlinks=true, citecolor=blue, urlcolor=blue, linkcolor=black}
\begin{document}
	\title{Target normal single-spin asymmetry in inclusive electron-nucleon scattering
		in the $1/N_c$ expansion  \footnote{Talk given at the ${\rm 25^{th}}$ International Symposium on Spin Physics: SPIN2023}}
	\author{J.~L.~Goity}
	\email[ E-mail: ]{goity@jlab.org}
	\affiliation{Department of Physics, Hampton University, Hampton, VA 23668, USA}
	\affiliation{Theory Center, Jefferson Lab, Newport News, VA 23606, USA}
	\author{C.~Weiss}
	\email[ E-mail: ]{weiss@jlab.org}
	\affiliation{Theory Center, Jefferson Lab, Newport News, VA 23606, USA}

\begin{abstract}
	The target normal single-spin asymmetry in electron nucleon scattering is studied in the framework of the $1/N_c$ expansion of QCD, which allows for a rigorous description in the energy range that includes the $\Delta$ resonance and below the second baryon resonance region. The asymmetry is driven by the absorptive part of the two-photon exchange component of the scattering amplitude, being therefore the most unambiguous two-photon exchange effect. Such amplitude is shown to be described up to the next to leading order in the $1/N_c$ expansion only in terms of the charge and magnetic form factors of the nucleons, consequence of the approximate $SU(4)$ spin flavor symmetry valid in the large $N_c$ limit for baryons.   A discussion is provided of the $1/N_c$ expansion framework along with the results for the asymmetries  in elastic ($e^- N^{\uparrow}\to e^- N$), inelastic ($e^- N^{\uparrow}\to e^- \Delta$), and inclusive scattering.      
	\end{abstract}
\maketitle
%

\section{Introduction}

Single spin-asymmetries are those asymmetries that result from effects in scattering where only one of the particles is polarized. There are different such asymmetries that can result from electroweak parity violation, such as is the case in inclusive scattering of longitudynally polarized electron on unpolarized target;   from correlations with more than one final momentum in semi-inclusive unpolarized electron scattering on a transversely polarized target; and from higher order absorptive effects, such as in inclusive unpolarized electron scattering on a transversely polarized target, which is the type of asymmetry discussed in this report. 

The target normal single-spin asymmetry (TSSA) in inclusive electron-nucleon scattering $e(k_{\rm i}) + N^\uparrow(p_{\rm i})
\rightarrow  e(k_{\rm f}) + X$, (Fig. 1) is defined by:
\begin{equation}
A_N=\frac{d\sigma_N}{d\sigma_U},
\end{equation}
where   the differential cross section is decomposed into the unpolarized $d\sigma_U$ and polarized $d\sigma_N$ components as follows:
\begin{equation}
{d\sigma} ={d\sigma_U}  + e_N^\mu a_\mu\;  {d\sigma_N} ,
\label{dsigma}
\end{equation}
where   $a^\mu$ is  the spin 4-vector of the target nucleon, and $e_N^\mu$ is the space-like  4-pseudovector given by:
$
e_N^\mu \equiv \frac{N^\mu}{\sqrt{-N^2}},
\hspace{1em}
N^\mu \equiv \epsilon^{\mu\alpha\beta\gamma} p_{{\rm i}\alpha} k_{{\rm i}\beta} k_{{\rm f}\gamma}$, which in the CM frame become: $e_N = (0, \bm{e}_N),
\hspace{1em}
\bm{e}_N = \frac{\bm{k}_{\rm i} \times \bm{k}_{\rm f}}{|\bm{k}_{\rm i} \times \bm{k}_{\rm f}|}$. The functions    $d\sigma_{U,N}$ depend only on the scattering angle $\theta$ in the CM frame, or the corresponding angle in the lab frame.

\begin{center}
\begin{figure}[h]
	\hspace*{0cm}	\includegraphics[width=.47\columnwidth]{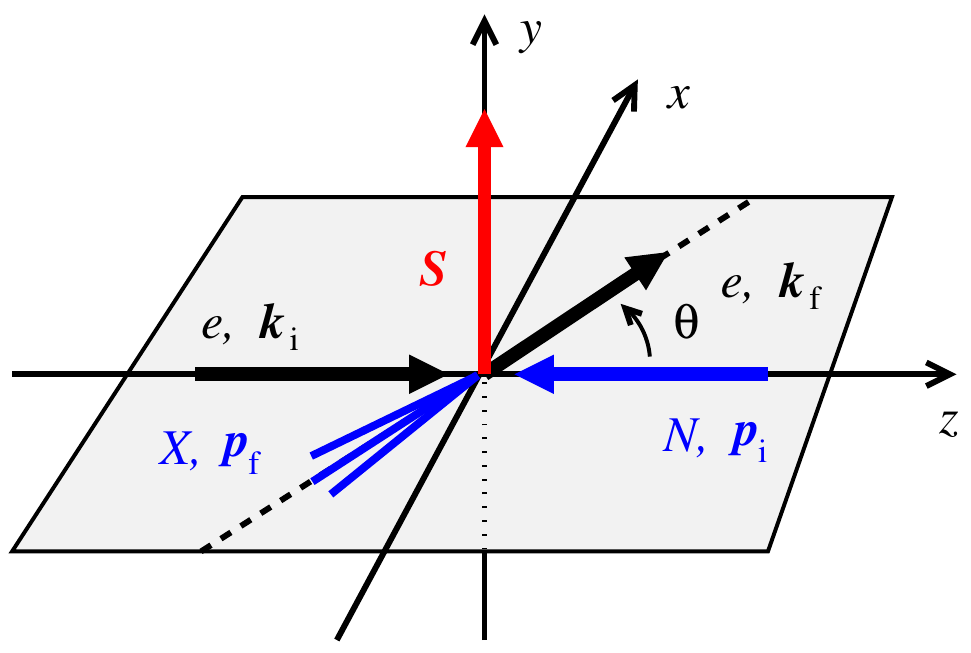}
	\caption[]{Inclusive electron-nucleon scattering in the electron-nucleon CM frame.
		The nucleon is polarized in the direction normal to the scattering plane.}
	\label{fig:cmframe}
\end{figure}
\end{center}

Experimentally, the asymmetry is then observed as a left-right cross section difference with respect to the plane defined by the incoming electron's momentum and the normal  polarization vector of the target nucleon. Such measurements were carried out long ago \cite{experiments} with inconclusive results due to the large error bars, and conclusive results are still to be realized. This is an important present challenge.

As a consequence of parity and time reversal invariance, the normal TSSA in inclusive electron scattering is necessarily a higher order QED effect, namely due to two-photon exchange. Moreover, it is entirely due to the absorptive component of the two-photon exchange scattering amplitude \cite{Barut-Fronsdal-Lee}. Thus the asymmetry results from the interference of the one-photon exchange with the absorptive part of the two-photon exchange amplitudes (Fig 2). The absorptive part is given by the product of the on-shell single photon electro-production amplitudes of the intermediate and final hadronic states and can be computed in this way if the amplitudes are known.
However there are clear  limitations, as among such amplitudes are those of electro-production on resonances, inaccessible in practice. Since the hadronic states that contribute to the absorptive two-photon exchange amplitude are only those kinematically allowed, in a regime of energy sufficiently low one can have   good  theoretical control, and also one can make efficient use of the said electro-production amplitudes.

\begin{figure}[h]
	\hspace*{.5cm}	\includegraphics[width=.48\columnwidth]{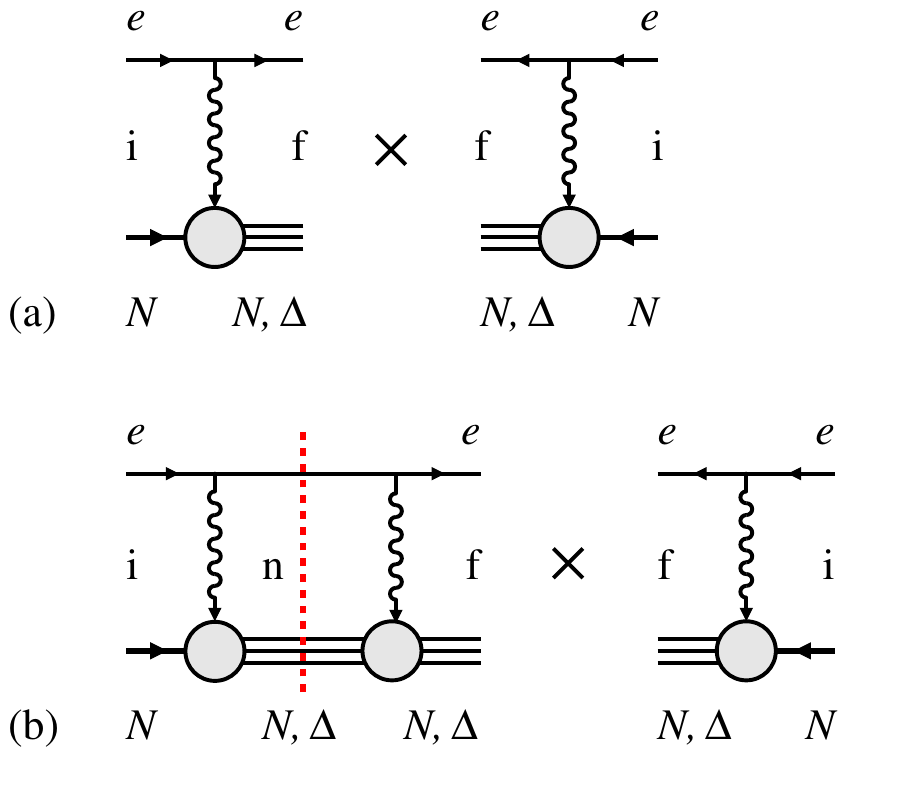}
	\caption[]{Inclusive electron-nucleon scattering cross section with $N$ and
		$\Delta$ final states in large-$N_c$ limit. (a) Spin-independent cross section from
		square of $e^2$ amplitudes. The circle denotes the electromagnetic current matrix element
		between baryon states. (b) Spin-dependent cross section from interference
		of $e^4$ and $e^2$ amplitudes. }
	\label{fig:diagrams}
\end{figure}

The discussion here will then be limited to the regime that includes the excitation of the $\Delta$ resonance and remains below the excitation of the higher resonances. In that regime one can make use of a powerful framework based on the $1/N_c$ expansion of QCD. In that expansion the baryon sector develops a dynamical spin-flavor symmetry $SU(2 N_f)$, $N_f=2$ or 3, the number of active light flavors. Consequence of that symmetry is that baryons belong into approximately degenerate spin-flavor multiplets where mass splittings between neighboring spin states with spins $\ord{N_c^0}$ are $\ord{N_c^{-1}}$, and QCD observables will therefore be represented as effective operators on the spin-flavor baryon multiplets. In the present case with only two flavors, the multiplet of interest is the one containing the nucleons, and is the symmetric $SU(4)$ multiplet that for generic $N_c$ contains the states $S=I=\frac 12, \cdots, \frac {N_c} 2$, i.e, $N$ and $\Delta$ for $N_c=3$. $SU(4)$ has 15 generators, namely the spin $S^i$, isospin $I^a$ and spin-flavor $G^{ia}$. The latter have matrix elements of $\ord{N_c}$ between states with spin $\ord{N_c^0}$, and in particular these operators, up to a multiplicative constant, represent at the leading order (LO) in a $1/N_c$ expansion several key couplings, namely, pion couplings to baryons, axial currents, and the magnetic component of the EM current. These are therefore enhanced in the large $N_c$ limit. In the context of implementing a $1/N_c$ expansion it is also necessary to include the baryon masses which scale as quantities $\ord{N_c}$. This naturally leads to making use of a non-relativistic expansion in all processes where the momenta involved are $\ord{N_c^0}$. With those general statements one can proceed to implement the $1/N_c$ expansion for the TSSA.

\section{Implementation of the TSSA in the $1/N_c$ expansion}

The EM current is implemented at arbitary $N_c$ making use of the assignment of quark charges $Q_q=\frac 3{2N_c}+I_3$, which is consistent with the Standard Model at arbitrary $N_c$ and in addition gives the ordinary charge formula for baryons, namely: $Q=1/2+\hat{I}^3$.
  The isosinglet (S) and isotriplet (V) components of the EM current at leading order in the the $1/N_c$ expansion, which automatically implies the non-relativistic expansion, read as follows \cite{FG} \footnote{  Terms in the currents with higher powers of momenta have been neglected, such as the isovector contribution to the charge component of the current, which stems from a relativistic correction and is proportional to $\frac {1}{m_N \Lambda}  g^{\mu 0} \eps^{\mu i j k} q^i  p^j \hat{G}^{ka} $, where $q$ and $p$ are the momentum transfer to the current and the baryon momentum respectively. Such terms are suppressed except at the upper end of the energy domain considered here, and are sub-leading. The electric quadrupole component of the current that mediates $N-\Delta$ transitions is suppressed by a factor $1/N_c^2$ with respect to the leading term \cite{quadrupole}, and is thus irrelevant to the   present calculation.}:
\begin{align}
J_S^{\mu}(q) &=\frac e2( G_E^S(q^2) g^{\mu 0} - i\frac{  G_M^S(q^2)}{\Lambda} \epsilon^{0\mu i j} q^i \hat{S}^j)
\nonumber \\
J_V^{\mu a}(q) &= e(G_E^V(q^2) \hat{I}^a g^{\mu 0}- i\frac 65 \frac{  G_M^V(q^2)}{\Lambda} \epsilon^{0\mu i j} q^i \hat{G}^{ja})\nonumber\\
J_{\rm EM}^\mu(q)&=J_S^{\mu}(q)+J_V^{\mu 3}(q),
\label{EMcurrent}
\end{align}
where $\Lambda$ is a  generic mass scale $\ord{N_c^0}$ and  $G_{E,M}^S$ and $G_{E,M}^V$ are  isoscalar and isovector   form factors, which are $\ord{N_c^0}$, and where for the choice $\Lambda=938$ MeV corresponds to the physical nucleon form factors \footnote{For the sake of convenience in the calculations and without significant difference the $G_E$ form factor is taken to be equal to the corresponding $F_1$ rather than the Sachs form factor. }. The order in the $1/N_c$ expansion of the components of the currents is thus determined by the order of the matrix elements of the associated operators, thus the electric charge terms and the isosinglet magnetic term are $\ord{N_c^0}$ while the isotriplet magnetic term is $\ord{N_c}$ because $G^{ia}$ has matrix elements of that order. In addition, the next order correction to the latter is $\ord{N_c^{-1}}$, i.e., two orders in $1/N_c$ with respect to the leading one. Other terms of the currents that are suppressed by the non-relativistic expansion, such as the spatial components of the convection current and the time components of the spin currents are sub-sub-leading with respect to the order of the present calculations. Thus, the EM currents \ref{EMcurrent} provide a complete description for the nucleon and $\Delta$ EM properties to the order of interest. 
	
	There is one important additional point related to the momentum transfer factor of the spin currents. When the electron energy is up to the range of the $\Delta$-$N$ mass difference, since this is $\ord{N_c^{-1}}$, such a momentum factor must be considered a quantity of that order. The correct way to organize the expansion in that regime   requires to impose a common power counting for the combined low energy and $1/N_c$ expansions, as it has been discussed in the case of its implementation in Chiral Perturbation Theory (e.g., \cite{FG}).  In the case of the proton for that range of momenta the electric charge term and the magnetic isotriplet are of the same order. 
	The range of momenta above the $\Delta$ excitation are on the other hand counted as quantities $\ord{N_c^0}$, and thus the leading term in the current is the isotriplet spin current. The calculations carried out here consistently treats both regimes, as it keeps all sub-leading contributions throughout.

\subsection{Calculation of the TSSA}
The interference term between the one- and two-photon exchange that contributes to the TSSA is given by:
\begin{align}
\sum_n{ {M}}^{(e2)^*}_{\rm  fi}{ {M}}^{(e4)}_{\rm fi}(B_{\rm n})\arrowvert_{Abs}+cc&= \sum_ne^6\frac{m_{B_{\rm n}}}{32 \pi^2\, t\sqrt{s}\;k_{\rm i} k_{\rm f}  k_{\rm n}}\nonumber\\
&\times {\rm Im}\left(\int d\Omega_{  \bm {k}_{\rm n}} \right. \left.\frac{ L_{\mu\nu\rho}(k_{\rm i},k_{\rm f},k_{\rm n}) H_{{\rm fi,n}}^{\mu\nu\rho}(k_{\rm i},k_{\rm f},k_{\rm n})}{(1-     \bm{\hat k}_{\rm i}\cdot  \bm{\hat   k}_{\rm n})(1-\bm{\hat k}_{\rm f}\cdot { \bm{\hat    k}_{\rm n}})}\right),
\label{interfterm}
\end{align}
where,$n$ indicates the baryon intermediate state in the box diagram ($N$ or $\Delta$), $k_{i,f,n}$ the CM momentum of the initial, final and box diagram electron,  and the leptonic and hadronic tensors are given by:
\begin{align}
L^{\mu\nu\rho}( {k}_{\rm i}, {k}_{\rm f}, {k_{\rm n}})&=Tr(\sh{\!k}_{\rm i} \gamma^\mu \sh{{\!k}_{\rm f}} \gamma^\nu \sh{\!k_{\rm n}}\gamma^\rho)\nonumber\\
H_{{\rm fi},{\rm n}}^{\mu\nu\rho}( {k}_{\rm i}, {k}_{\rm f}, {k_{\rm n}})&=\langle B_{\rm i}\mid (J_{\rm EM}^\mu(k_{\rm i}-k_{\rm f}))^\dagger\mid B_{\rm f}\rangle\nonumber\\
&\times\langle B_{\rm f}\mid J_{\rm EM}^\nu(k_{\rm n}-k_{\rm f})\mid B_{\rm n}\rangle \langle B_{\rm n}\mid J_{\rm EM}^\rho(k_{\rm i}-k_{\rm n})\mid B_{\rm i}\rangle .
\label{Hadronic tensor}
\end{align}
The current matrix elements between the baryon states are evaluated with the $1/N_c$ expanded current operators using dipole form  factors, where the phase space integrations can be carried out in close analytic form. The different contributions in \ref{interfterm} can be organized in terms of t-channel angular momentum and isospin quantum numbers for both the box and total contributions. Such decompositions show a hierarchy in $1/N_c$ powers, namely with the $I=J$  contributions being the dominant ones. In addition, and upon summing the different contributions required by gauge invariance,  for each separate set of possible contributions of the components of the EM currents and for each separate t-channel angular momentum and isospin one finds that such contributions to the interference cross section are free of infrared and collinear divergencies. This also holds for each intermediate state in the box and final state $N$ or $\Delta$.

\subsubsection{Large $N_c$ limit}
As an interesting first step one can consider the large $N_c$ limit. In this case only the isotriplet component of the spin current is relevant, and the $N$ and $\Delta$ are degenerate. This calculation \cite{GWW1} gives some important insights that carry on to the next to leading order (NLO) calculation as shown below. In that calculation one observes that the contribution of the $\Delta$ as intermediate state is very important for the elastic asymmetry (nucleon final state). On the other hand, the inelastic asymmetry ($\Delta$ final state) is very small if the momentum dependence of the form factors is neglected.  However, the inclusion of the form factors leads to an important qualitative change where   the inelastic asymmetry becomes important as well.  

\subsection{TSSA to the NLO }
Here a summary of the full calculation \cite{GWW2} is presented. The different contributions to the TSSA, namely elastic ($N$ final state), inelastic ($\Delta$ final state) and inclusive are discussed. First, the case where the form factor's dependencies are neglected. In this case results for the TSSA can be given in a relatively compact form showing explicitly  the $N_c$ dependencies, with obvious notations:
\begin{align}
\frac{d\sigma_N}{d\Omega}  (N_{\rm i},N,N) &= \frac{ \alpha ^3 k^2 m_N^3 }{4000\, \Lambda ^3\,
	s^{3/2}\, t \,(1+x)} \Big(2 (1+x)-(x+3) \log  \frac{1-x}{2}\Big)\nonumber\\
&\times
\big((N_c-3) G_M(-I_3)-(N_c+7) G_M(I_3)\big)^2 \nonumber\\
&\times\big(10
\Lambda  G_E(I_3)+k ((N_c-3)
G_M(-I_3)-(N_c+7) G_M(I_3))\big) \nonumber\\
\frac{d\sigma_N}{d\Omega}  (N_{\rm i},N,\Delta) &= \frac{\Theta (k_\Delta)\alpha ^3 m_N^2 m_\Delta }{2000 \,\Lambda ^3\, s^{3/2}\, t\, (1+x)}  (N_c-1) (N_c+5)
(G_M(-I_3)-G_M(I_3))^2 \nonumber\\
&\times \Big(2
k k_\Delta\, (1+x)-\log  \frac{1-x}{2}\;\, \big(k^2 (1+x)+2
k_\Delta^2\big)\Big) \nonumber\\
&\times\Big(k ((N_c-3)
G_M(-I_3)-(N_c+7) G_M(I_3))-5 \Lambda 
G_E(I_3)\Big) \nonumber
\\
\frac{d\sigma_N}{d\Omega}  (N_{\rm i},\Delta,N) &=\frac{\Theta (k_\Delta)\alpha ^3 k_\Delta m_N^2 m_\Delta }{16000\, \Lambda ^3 \,s^{3/2}\, t \,(1+x)}  (N_c-1) (N_c+5)
(G_M(-I_3)-G_M(I_3))^2   \nonumber\\
&\times\Big(2
\log \frac{1-x}{2}\;\, \Big(20 \Lambda  G_E(I_3) (2
k-k_\Delta (1+x))-((N_c-3) G_M(-I_3)\nonumber\\
&-(N_c+7)
G_M(I_3)) \Big(2 k^2-3 k k_\Delta\, (x-1)-2  k_\Delta
^2 (x-2)\Big)\Big)\nonumber\\
&-(1+x) \Big(\Big(11 k^2-k k_\Delta\,+4
k_\Delta^2\Big) ((N_c-3) G_M(-I_3)-(N_c+7)
G_M(I_3)) \nonumber\\
&-40 \,\Lambda \, G_E(I_3) (k- k_\Delta
)\Big)\Big)  \nonumber
\end{align}\begin{align} 
\frac{d\sigma_N}{d\Omega}  (N_{\rm i},\Delta,\Delta) &=  \frac{\Theta (k_\Delta)\alpha ^3 k_\Delta^2 m_N^2 m_\Delta }{80000\, k \,\Lambda ^3\, s^{3/2} \,t \,(1+x)}  (N_c-1) (N_c+5)
(G_M(-I_3)-G_M(I_3))^2  \nonumber\\
&\times
\Big(200 \,\Lambda \, G_E(I_3) \big((1+x) (k-k_\Delta)+\log
\frac{1-x}{2}\;\, (k (1+x)-2 k_\Delta)\big)\nonumber\\
&+((N_c-23)
G_M(-I_3)-(N_c+27) G_M(I_3)) \big(2 \log
\frac{1-x}{2}\nonumber\\
&\times \big(-6 k^2+k k_\Delta\, (5 x+3)-6
k_\Delta^2\big)+k_\Delta (1+x) (9 k-23 k_\Delta
)\big)\Big)  ,\label{dsigmaSSA}
\end{align}
where $(N_i,B,B_n)$ represents the  initial (target nucleon with isospin $I_3$), the final $B=N,\;\Delta$ and the baryon in the box $B_n=N,\;\Delta$.    $G_{M,E}(I_3)$ are the form factors neglecting their momentum transfer dependency, and $s$, $t$ are the Mandelstam invariants, 
$x=\cos\theta$,  and
$t=-2 k_{\rm i} k_{\rm f} (1-x)$. For convenience in the following the  definition of the TSSA is taken to be $A_N=d\sigma_N/d\sigma_{\rm elastic}$. One immediately checks that at LO $A_N=\ord{\alpha N_c}$. The LO contributions stem from the $I=J$ t-channel exchange of the two-photon amplitude driven by the isotriplet spin current. The sub-leading contributions result from the sub-leading terms in the EM current and $I\neq J$ exchanges of the isotriplet spin current.  The $N_c$ dependence shown in Eq.\ref{dsigmaSSA} results from the matrix elements of the spin-flavor operators \cite{GWW2}.

\begin{center}
	\begin{figure}[h]
		\includegraphics[width=5cm,height=3.48cm]{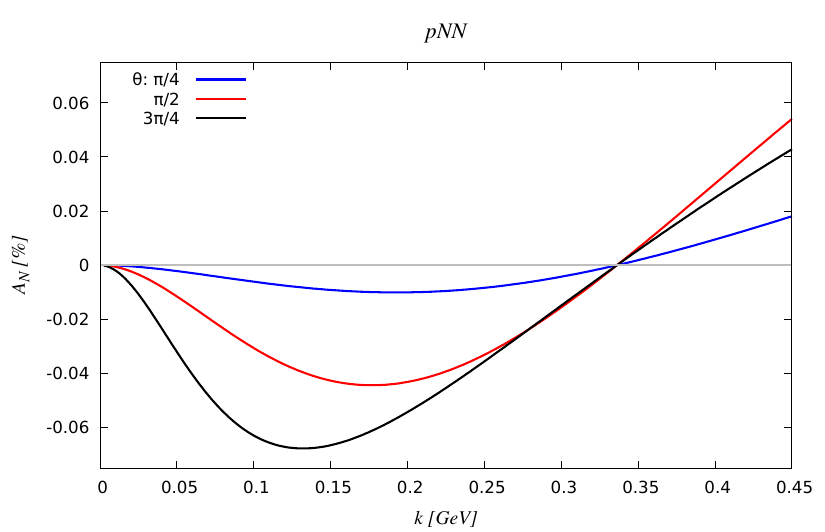}\;\includegraphics[width=5cm,height=3.48cm]{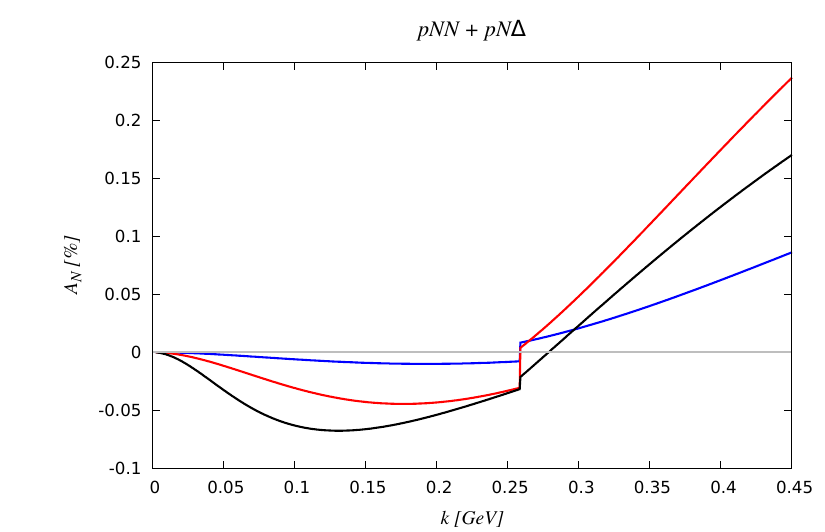}\;\;	\includegraphics[width=5cm,height=3.48cm]{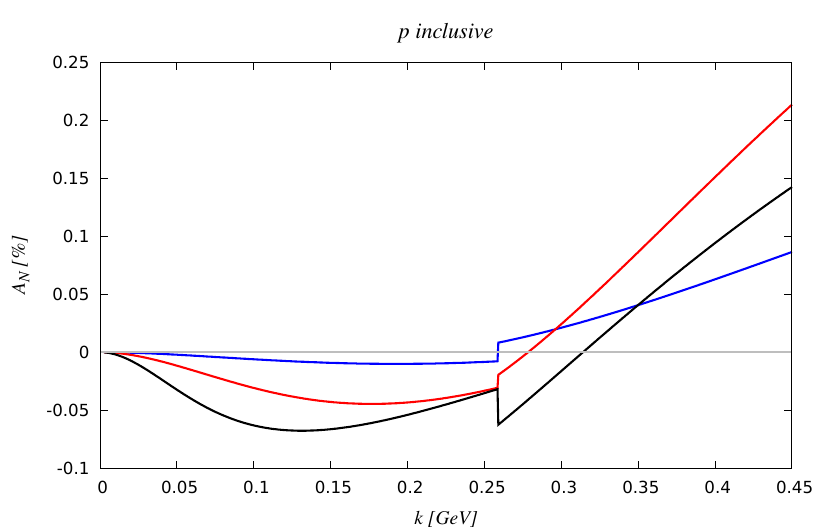}\\ \includegraphics[width=5cm,height=3.48cm]{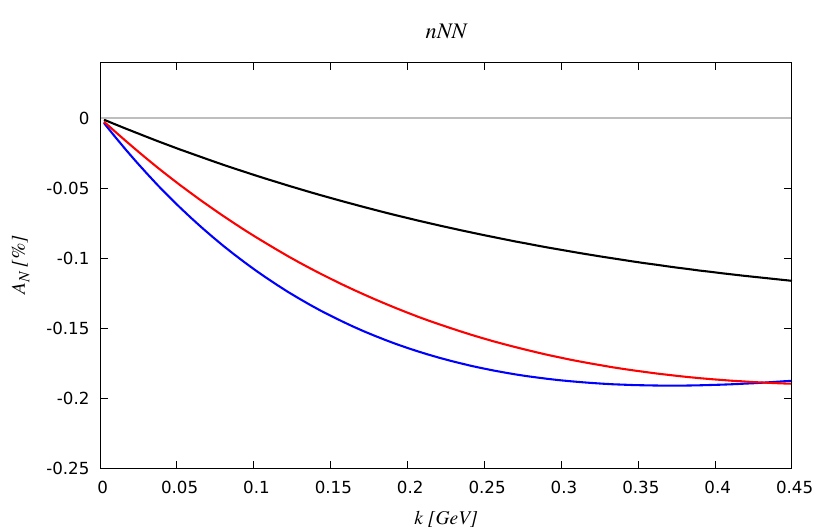} \;\; \includegraphics[width=5cm,height=3.48cm]{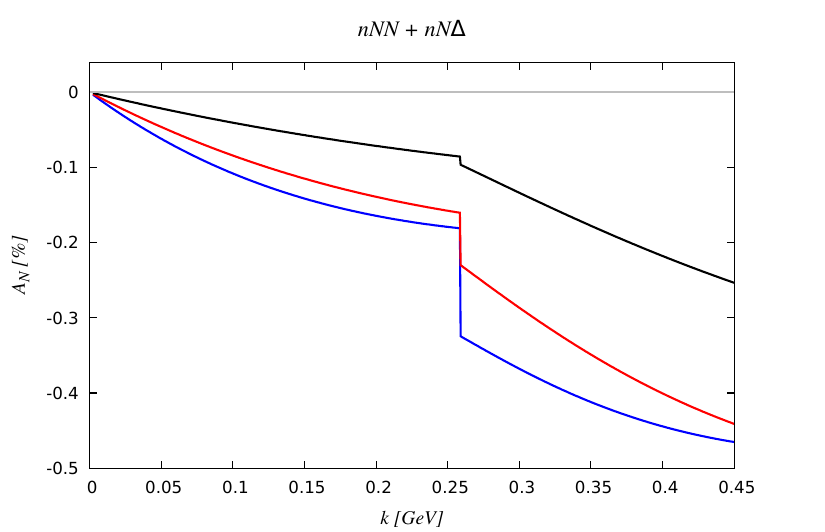}
		\includegraphics[width=5cm,height=3.48cm]{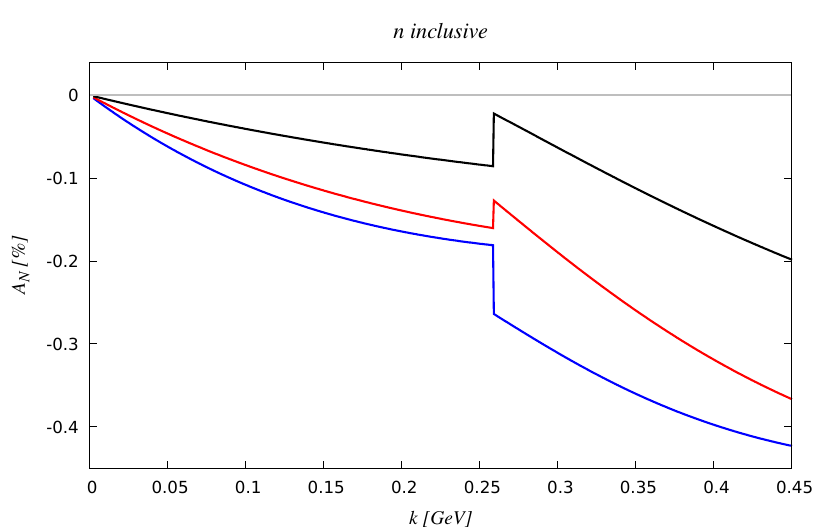}
		\caption[]{  $A_N$ vs $k$ with no form factors and stable $\Delta$. Top (bottom) panels proton (neutron). Left: elastic  with only nucleon in the TPE amplitude, middle: elastic  with nucleon and $\Delta$ in TPE amplitude, right:     inclusive.  }
		\label{fig:TSSAElastic}
	\end{figure}
\end{center}
The TSSAs   results are shown in Fig. 3. In the LO the proton and neutron asymmetries would be equal with opposite sign. At low energy the asymmetry in the proton is affected by its electric charge (left panels); at low energy the magnetic terms have a momentum suppression and thus the electric term of the current gains significance. The elastic asymmetries receive an important enhancement from the $\Delta$ contribution to the two-photon exchange amplitude. Note the discontinuity in the asymmetry at the   $\Delta$ threshold: it results from the  contribution in the box diagram in the domain where the two photons tend to be real and collinear. Such a discontinuity is smoothed out when the actual decay width of the $\Delta$ is taken into account. In the absence of $t$-dependence for the form factors there is a fine cancellation that suppresses the inelastic asymmetry ($\Delta$ in final state). This is what was observed initially in the purely LO calculation \cite{GWW1} mentioned earlier.

\begin{center}
	\begin{figure}[t]
		\includegraphics[width=6cm,height=4.8cm]{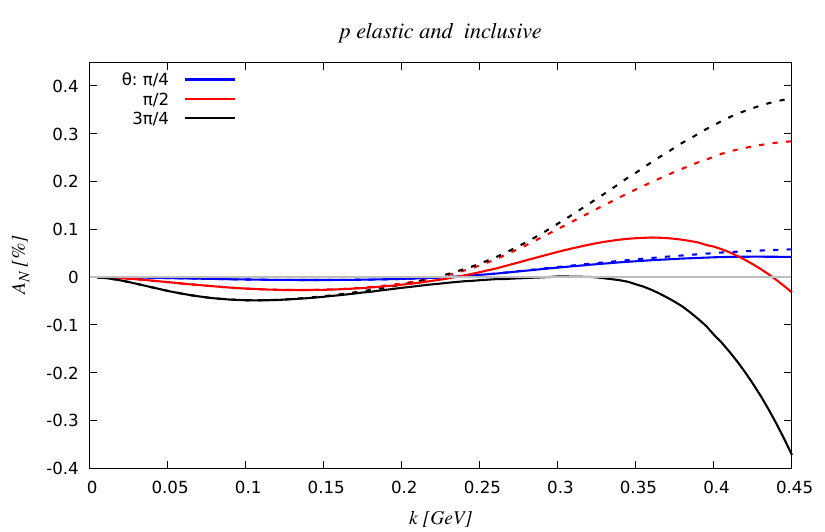} \includegraphics[width=6cm,height=4.8cm]{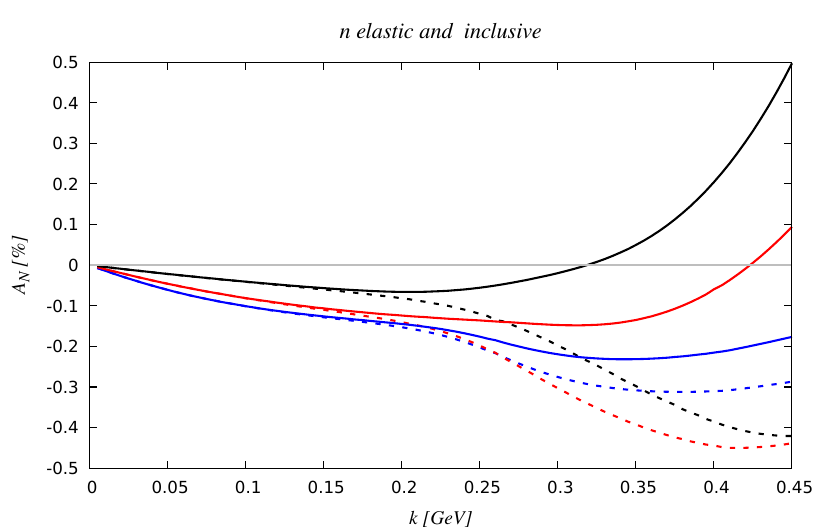}\\	\includegraphics[width=6cm,height=4.8cm]{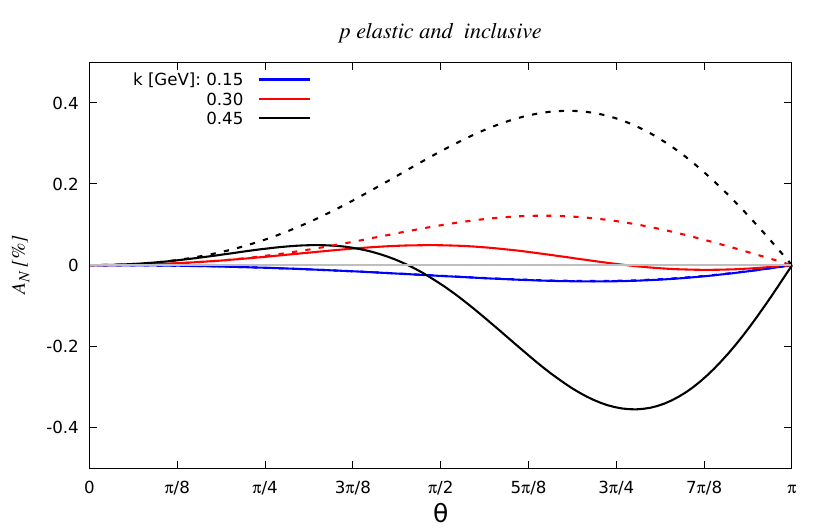}\includegraphics[width=6cm,height=4.8cm]{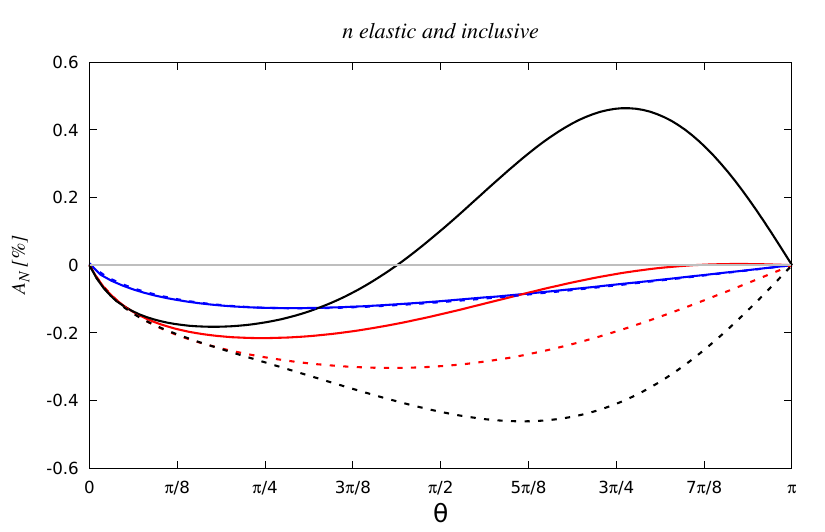}
		\caption[]{  $A_{N}$ vs $k$ (top) and $A_{N}$ vs $\theta$ (bottom) with inclusion of form factors and $\Delta$ width.  Elastic (dashed) and inclusive (solid). }
		\label{fig:TSSAFF}
	\end{figure}
\end{center}

 In Fig. 4 the results in the realistic case with form factors and $\Delta$ decay width are shown. The form factors are described by a common dipole form $\Lambda_{EM}^4/(Q^2+\Lambda_{EM}^2)^2$ with $\Lambda_{EM}^2=0.71 {\rm ~ GeV}^2$, and the effect of the $\Delta$ decay width is implemented in the Breit-Wigner approximation with $\Gamma_\Delta=0.125$ GeV. These effects preserve the property that the TSSA is free of infrared and collinear divergencies.
 
 The form factors are of key importance, manifested in the dramatic enhancement of the inelastic asymmetry. In fact, at CM scattering angles $\theta> \pi/2$ the inelastic asymmetry, which has opposite sign to the elastic, dominates in the inclusive asymmetry.  The previously mentioned fine cancellations that suppressed the inelastic asymmetry are disrupted by the form factors, and thus the large effect. Importantly, it is the LO contributions that give rise to this effect. Since both the elastic and inclusive asymmetries can in principle be separately measured, these predictions can be put to the test.
 
 It is useful to compare the LO and NLO results shown in Fig. 5. For the neutron the NLO effects are small, while one sees significant effects in the proton. The reason for the latter is that the   charge term in the current plays an important role, the reason being that, although NLO in $1/N_c$, it is not suppressed by a momentum factor as is the case of the spin component of the current as discussed earlier. This results in an important contribution at low momenta and which remains significant through the onset of the $\Delta$ contributions. 
 
 \begin{center}
 	\begin{figure}[h]
 		\includegraphics[width=6cm,height=4.8cm]{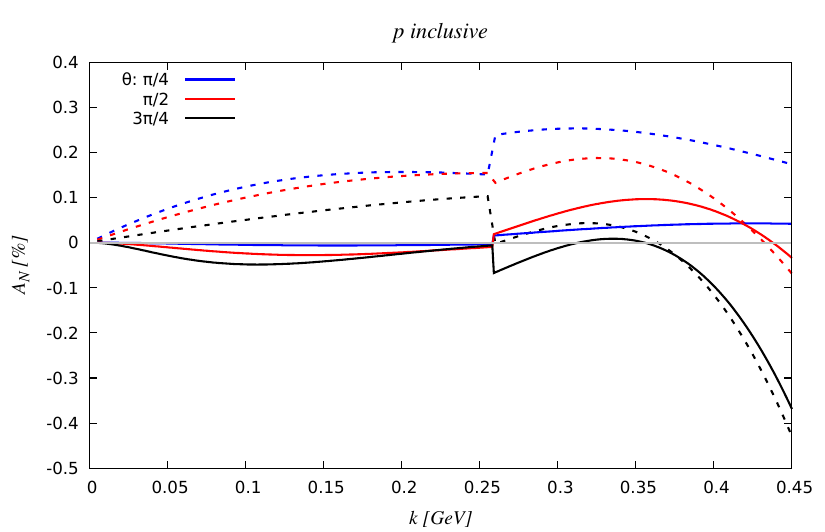} \includegraphics[width=6cm,height=4.8cm]{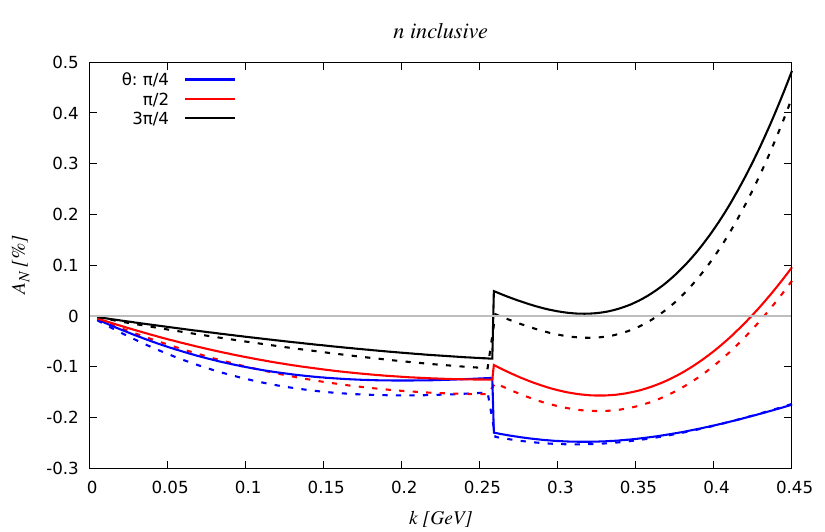}\\	
 		\caption[]{Inclusive $A_{N}$ for proton (left panel) and neutron target (right panel). Comparison between the LO
 			in $1/N_c$ (dashed lines) and the results of this work (solid lines). Form factors are included, and the physical
 			phase space is used in the LO result.}
 		\label{fig:TSSA-LO-vs-NLO}
 	\end{figure}
 \end{center}

 The magnitude of the TSSA is in the $10^{-3}-10^{-2}$ range. While   asymmetries of this magnitude could   not be measured as nonzero in the old experiments \cite{experiments}, today it would be relatively straightforward to do so. What is needed is a polarized target and an electron beam with the energy range of the present discussion, such as the  MAMI facility in Mainz \cite{MAMI}. 
 
 One contribution to the asymmetry that was left out is the continuum one that is dominated by the pion-nucleon continuum. The inclusion of the $\Delta$ width has taken care of the resonant $\pi N$ piece, and the rest of the contributions  are actually suppressed at low energy by chiral symmetry (extra two powers of momentum) and in the $N_c$ power counting they are $\ord{N_c^{-1}}$ with respect to the leading order. Such contributions can compete with the sub-leading contributions just calculated only in the upper energy range considered. It is expected that they will have almost negligible effect on the asymmetries.

 Attempting to extend the framework in a consistent fashion to higher energies, which requires the inclusion of higher $N$ and $\Delta$ resonances is  challenging. One can however point out that individual resonances will contribute at $\ord{N_c^{-1}}$ with respect to the LO considered here. The reason is that the matrix elements of the EM current between a resonance and a ground state baryon receives a suppression factor $1/\sqrt{N_c}$ \cite{GS}. In this case an approach based on dispersion theory is at present the most accessible one in spite of the limited  experimental information that is  insufficient for a complete   determination of the asymmetry (e.g. \cite{PVdH}). At high energies the TSSA is studied in the DIS regime where it can be treated along lines such as in Ref. \cite{ASW}.
 
 \section{ Summary and comments}

 The TSSA in electron nucleon scattering was studied in this new application of the $1/N_c$ expansion of QCD. The approach is well constrained in terms of the nucleon's form factors thanks to the $SU(4)$ symmetry of the large $N_c$ limit. It was shown that up to the NLO in the expansion in $1/N_c$ the TSSA can be unambiguously predicted in the energy range that includes the $\Delta$ resonance and below the onset of the higher resonances. The expansion permits for a hierarchical organization of the different contributions to the asymmetry. Some important conclusions have been drawn, in particular the important role of the $\Delta$ and the sensitivity of the TSSA to the form factors. The authors expect that the results obtained here are solid predictions for this most directly observable effect of two-photon exchange. Experiments aimed at measuring the TSSA, if possible the elastic and inelastic ones separately, would therefore be very   informative.
 
 Two-photon exchange effects in electron-nucleon scattering remain an important area of interest, both experimental and theoretical. Such effects are  relevant for elastic electron  scattering in  the extraction of the EM form factors, where they have different weight in the Rosenbluth separation and the polarization transfer methods   as   manifested in the extraction of the ratio $G_E/G_M$ of the proton. The two-photon exchange contribution to the  elastic scattering amplitude can be accessed more directly by comparison of electron vs. positron scattering, which has recently motivated the proposal of such measurements at Jefferson Lab \cite{Grauvogel} and DESY \cite{DESY} (all of these in a higher range of energy than covered by the present work). The TSSA represents in reality the most pristine two-photon exchange effect, as it vanishes in the one-photon exchange approximation and is entirely given by absorptive effects which allow for a more directly accessible theoretical analysis.
 
 \section{Acknowledgments}

This material is based upon work supported by the U.S.~Department of Energy, Office of Science,
Office of Nuclear Physics under contract DE-AC05-06OR23177 (JLG and CW), and by the National Science Foundation, Grant Number PHY 1913562 (JLG).

%

\end{document}